\documentclass[%
 ,twocolumn%
 ,secnumarabic%
,amssymb, amsmath,nobibnotes, aps, prl]{revtex4}
\usepackage{docs}%
\usepackage{bm}
\usepackage{graphicx}
\expandafter\ifx\csname package@font\endcsname\relax\else
 \expandafter\expandafter
 \expandafter\usepackage
 \expandafter\expandafter
 \expandafter{\csname package@font\endcsname}%
\fi

\begin{document}

\title{A Framework for Creating Galaxy Models in the Geometry of the Conservation Group with Dark Matter Halos and Flat Rotation Curves}%

\author{Edward Lee Green}%
\email{egreen@ung.edu}
\affiliation{University of North Georgia, Dahlonega, Georgia 30597}
\date{October 2022}%

\begin{abstract}
Pandres has developed a theory which extends the geometrical structure of a real four-dimensional space-time via a field of orthonormal tetrads with an enlarged covariance group.  This new group, called the conservation group, contains the group of diffeomorphisms as a proper subgroup.  The free-field Lagrangian density is $C^\mu C_\mu \sqrt{-g}\,$, where $C^\mu$ is a vector which measures curvature.  When massive objects are present a source term is added to this Lagrangian density.   The weak-field approximation implies that gravitational waves travel at the speed of light.  Spherically symmetric solutions for both the free field and the field with sources are found.  In the free-field case, the field equations require nonzero stress-energy tensors.  However, we find that for our model to be an acceptable model, we must have a source term in the Lagrangian.  In our framework we divide up the galaxy into three spherically symmetric regions: a baryonic matter-dominated central bulge, a dark matter-dominated mesosphere and an outside region where neither type dominates.    Assuming the density of baryonic matter has a central cusp, we show how to model the bulge.  Via an isothermal condition we find a model for the mesosphere and show this model implies flat rotation curves with one free parameter.  The outside region is readily modeled via previously published results. The models for the bulge, mesosphere and outside region are combined into one continuous model.  Using the radial acceleration relation we then show how a galaxy model may be set up for a rotationally supported galaxy.  

Orchid: 0000-0002-8193-4109
\keywords{ Extended Covariance Group;  Spherical Symmetric Solutions; Galaxy Rotation Curves}

\end{abstract}

\maketitle

\tableofcontents

\section{Introduction.}

Let ${\cal X}^4$ be a 4-dimensional space where the field variables are the orthonormal tetrad $h^{i}_{\,\,\mu}$ \cite{Green2020a,Green2020,Green2009} .  We assume the field variables are differentiable on ${\cal X}^4$ and have nonvanishing determinant.   We use natural units with $G=c=1$.  Using $h^i_{\;\mu}$, a metric $g_{\mu \nu}$ may be defined on ${\cal X}^4$ by $g_{\mu\nu}=\eta_{ij}\, h^i_{\,\,\mu}h^j_{\,\,\nu}$ where $\eta_{ij} = diag\bigl\{-1,1,1,1\bigr\}$ and the space may be interpreted as a Riemannian manifold.   We use the corresponding metric to define the usual covariant derivative, denoted by use of the semicolon.  

Einstein extended special relativity to general relativity by extending the group of transformations from the Lorentz group to the group of diffeomorphisms \cite{Einstein1949}.  We  \cite{Pandres1981,Pandres1984,Pandres2009,Green2009,PG2003,Green2020} further extend the covariance group by finding the largest group of transformations where a conservation law of the form $V^\alpha_{\; ;\alpha} = 0\, $ is invariant.   For a transformation from coordinates $x^\nu \to x^{\overline{\alpha}}$, this condition has been shown to imply \cite{Pandres1981}
\begin{equation}x^\nu_{\;\; ,\overline{\alpha}}\bigl(x^{\overline{\alpha}}_{\;\; ,\nu ,\mu} -
x^{\overline{\alpha}}_{\;\; ,\mu , \nu} \bigr) = 0 \quad . \label{e1} \end{equation}
The group of transformations which satisfy (\ref{e1}) is called the {\it group of conservative transformations} and we easily see it contains the group of diffeomorphisms as a proper subgroup.    Since the scalar wave equation may be written as $\Psi^{,\alpha}_{\;\; ;\alpha} = 0$ , we see that the conservation group is "the largest group of coordinate transformations under which the scalar wave equation is invariant"\cite{Pandres2009}.  The conservation group shows potential for unifying the fields of nature \cite{Pandres2009}.

Analogous to the Riemann curvature tensor, we have the {\it curvature vector} defined by 
\begin{equation} C_\alpha \equiv h_i^{\,\,\nu}\bigl(h^i_{\,\, \alpha ,\nu}-h^i_{\,\, \nu ,\alpha} \bigr) \label{Cmu}
\end{equation}
Using the Ricci rotation coefficients, $\gamma^i_{\;\;\mu\nu}=h^i_{\;\mu ;\nu}$, we may also see that $C_\alpha = \gamma^\mu_{\;\;\alpha\mu}$ \cite{Pandres2009}.   Pandres has shown \cite{Pandres1981} that $C_\alpha$ is covariant under transformations from $x^\mu$ to $x^{\overline{\mu}}$ if and only if the transformation is conservative and thus satisfies (\ref{e1}).  Furthermore \cite{Pandres2009}, there exists a conservative transformation that will convert the orthonormal tetrad $h^i_{\;\mu}$ into constants (i.e. a ``flat space'') if and only if $C_\alpha = 0$.

We will assume that $h^i_{\;\mu}$ is a function of position that makes it possible to interpret a particular solution of our field equations as a manifold with metric $g_{\mu\nu} = \eta_{ij}h^i_{\;\mu}h^j_{\;\nu}$.  We will call this the {\it associated manifold.}  Classically the tetrad is used to represent a local observer.   It has been shown that  the {\it conservative transformation group}  has properties that would be expected if observers are not classical observers but quantum observers \cite{Pandres1984}.  We note that if $x^{\bar{\mu}}_{\; ,\nu}$ is a non-diffeomorphic but conservative transformation, then in the {\it associated manifold}, $x^{\bar{\mu}}$ may be interpreted as anholonomic coordinates for the first observer \cite{Schouten1954} whose coordinates are  $x^{\mu}$.   Another interpretation that recognizes the larger geometry is that the conservative, nondiffeomorphic transformations from $x^\mu$ to $x^{\bar{\mu}}$ is a transformation from one manifold to a second manifold, since a non-diffeomorphic transformation would alter the Riemannian curvature tensor,  $R^\alpha_{\;\;\beta\mu\nu}$.  These will be called the {\it Conservative Family} of {\it associated manifolds.}  Previous results suggest that $C^\mu C_\mu$ is related to mass-energy which would imply that every {\it associated manifold} in the {\it Conservative Family} has the same mass-energy.  We speculate that the conservative group, which is similar in mass-energy conserving properties to the unitary transformation group used in quantum theory, may be the foundation for a quantum theory of gravity.

Historically, the use of a stationary coordinate system on the surface of the earth resulted in the introduction of a ficticious force (Coriolis force).   Yet such a faulty coordinate system is natural and also useful in many situations.  Similarly we naturally and usefully interpret our world as a manifold, particularly the {\it associated manifold}.   Some forces which are actually ficticious are introduced by this viewpoint (such as perhaps dark matter and dark energy).   In our perceived Riemannian manifold, the geometry and physics are determined by the standard tensors of General Relativity.   Based on these assumptions, we will investigate the geometry of the physically implied {\it associated manifold} and attempt to find a reasonable physical model.  

\subsection{The Free Field Lagrangian.}
 The simplest scalar constructed from $C_\alpha$ is $C^\alpha C_\alpha$ which is used to define the Lagrangian for the free field:  
\begin{equation} {\cal L}_f= \frac1{16\pi}\int C^\alpha C_\alpha \, h \; d^4 x  \label{e2} \end{equation} where $h = \sqrt{-g}$ is the determinant of the tetrad.
We may write our Lagrangian density in terms of  \cite{PG2003,Green2020} 
\begin{equation}  C^\alpha C_\alpha = R + \gamma^{\alpha \beta \nu} \gamma_{\alpha \nu \beta}-2C^\alpha_{\; ;\alpha}  \label{e4} \end{equation}
where $R$ is the usual Ricci scalar curvature and $\gamma^{\alpha}_{\;\beta\nu} = h_i^{\;\alpha}h^i_{\; \beta ; \nu}$ are the Ricci rotation coefficients.  Comparing (\ref{e4}) with GR we see that the Lagrangian density of the free field (\ref{e2}) contains additional terms which may model dark matter \cite{PG2003,Green2020}.  

Setting $\delta {\mathcal L}_f \, = 0$ leads to field equations and we vary only the usual orthonormal tetrad, $h^i_{\;\alpha} \, $.  In this case, the field equations \cite{Pandres1981,Pandres1984,Pandres2009,PG2003} are:
\begin{equation} C_{\mu ;\nu} - C_{\alpha}\gamma^\alpha_{\;\mu\nu} - g_{\mu\nu}C^\alpha_{\; ;\alpha} - \frac12 g_{\mu\nu}C^\alpha C_\alpha \; = \;
0  \label{fe0} \end{equation}
In the {\it associated manifold}, we also accept Einstein's equations $G_{\mu\nu} = 8\pi T_{\mu\nu}$.  We adopt the Schrodinger point of view:  ``I would rather you did not regard these equations as field equations, but as a  definition of $T_{ik}$, the matter tensor \cite{Schrodinger1950}.'' For the {\it associated manifold}, we use the identity for the Einstein tensor:
\begin{eqnarray}  G_{\mu\nu}= \; &  C_{\mu ; \nu}- C_\alpha \gamma^\alpha_{\; \mu\nu} -g_{\mu\nu}C^\alpha_{\; ;\alpha}-\frac12 g_{\mu\nu}C^\alpha C_\alpha   \nonumber \\ &  \quad  + \; \gamma^{\;\;\alpha}_{\mu \;\; \nu ;\alpha}+\gamma^\alpha_{\;\;\sigma
\nu} \gamma^\sigma_{\;\; \mu \alpha} + \frac12 g_{\mu\nu}\gamma^{\alpha \beta
\sigma} \gamma_{\alpha \sigma \beta}  \nonumber \end{eqnarray}   and  (\ref{fe0}) to see that the field equations imply that the free-field stress-energy tensor is  
\begin{equation}8\pi \bigl(\mathbf{T}_{\rm f}\bigr)_{\mu\nu} = \gamma^{\;\;\alpha}_{(\mu \;\; \nu) ;\alpha}+  \gamma^\alpha_{\;\;\sigma (\nu} \gamma^\sigma_{\;\; \mu) \alpha}  + \frac12 g_{\mu\nu}\gamma^{\alpha \beta
\sigma} \gamma_{\alpha \sigma \beta}\;    \label{e9} \end{equation}
This implies that when no sources are present in the ${\cal X}^4$ space, the field equations produce a stress-energy tensor which may be nonzero in the {\it associated manifold}.  This automatic source is due to the fact that the manifold is an approximation of the more general geometry of ${\cal X}^4$.   We speculate that these terms correspond to dark matter or dark energy and that instead of putting the dark sector into our theory in an ad hoc way, our interpretation of our world as the {\it associated manifold} implies that we will observe the dark sector.

\subsection{Total Lagrangian.}
When sources are present, the Lagrangian is of the form
\begin{equation} {\cal L}= {\cal L}_{\rm f}+{\cal L}_{\rm s} =  \int \biggl( \frac1{16\pi}C^\alpha C_\alpha + L_s \biggr) \, h \; d^4 x  \label{e10} \end{equation}  where $L_s$ is the appropriate source Lagrangian density function.  Using (\ref{e4}), one may write the density as
$\frac1{16\pi}\Bigl(R + \gamma^{\alpha \beta \nu} \gamma_{\alpha \nu \beta}-2C^\alpha_{\; ;\alpha} + 16\pi L_s \Bigr)
\; $
where $R$ is the usual Ricci scalar curvature.   As noted by \cite{Weinberg1972}, the coefficient of $h\, h^{i\nu}\delta h_i^{\;\mu}$ may be identified as the negative of the stress-energy tensor.   Thus using (\ref{fe0}), the full coefficient of $h\, h^{i\nu}\delta h_i^{\;\mu}$ is found to be
\begin{equation} \frac{1}{8\pi}\biggl(C_{\mu ;\nu}- C_\alpha \gamma^\alpha_{\; \mu\nu} - \frac12 g_{\mu\nu}C^\alpha C_\alpha - g_{\mu\nu}C^\alpha_{\; ;\alpha} \biggr)  - (T_{\rm s})_{\mu\nu}
\nonumber 
\end{equation}
Requiring this variation should be idetically zero leads to the source stress-energy tensor given by
\begin{equation}  8\pi (\mathbf{T}_{\rm s})_{\mu\nu} \; = \;   C_{(\mu ;\nu)}- C_\alpha \gamma^\alpha_{\; (\mu\nu)} -\frac12 g_{\mu\nu}C^\alpha C_\alpha - g_{\mu\nu}C^\alpha_{\; ;\alpha}  \label{e11} \end{equation}
with symmetrization indicated by the parentheses.  
Using $G_{\mu\nu} = 8\pi T_{\mu\nu}$ and (\ref{e9}) and (\ref{e11}) we also find the total stress-energy tensor to be
\begin{equation} 
\mathbf{T} \; = \; \mathbf{T}_{\rm f}  + \; \mathbf{T}_{\rm s}    \label{e12} \end{equation}
In other words, the total stress-energy is the direct sum of the free-field stress energy tensor and the source stress-energy tensor.  The free-field part corresponds to the dark matter/dark energy  and the source part corresponds to ordinary matter and fields.  The source stress-energy term (\ref{e11}) transforms covariantly under the conservation group (\ref{e1}).   By the way, as in General Relativity, we note that individually these are {\it not} covariant divergenceless, but the total stress energy does have the property that $T^\mu_{\; \nu ; \mu} = 0$.  

We propose that in order to model a physical system, one should determine the appropriate structure of $\mathbf{T}_{\rm s}$ in a suitable coordinate system and then find a tetrad solution of (\ref{e11}).  Secondly, identify and determine acceptable solutions for $\mathbf{T}_{\rm f}$.  

In the static case (no dependence of the field variables on $x^0$ which represents time) we find that $C_{0\, ; 0} - C_\alpha \gamma^\alpha_{(0 \, 0)} =0$.  Now the appropriate source Lagrangian density for a perfect fluid is $\rho_s$, the energy density \cite{dFC1990}.  It is also well known that adding a pure divergence to the Lagrangian density does not affect the field equations.  Thus in the static case with the $C^\alpha_{\; ;\alpha}$ term absorbed into the density, we find from (\ref{e11}) that the density of mass-energy of the source, $\rho_s$, is given by $8\pi \rho_s \; \equiv \; \frac12\, C^\alpha C_\alpha$ or
\begin{equation} \rho_s \; \equiv \; \frac1{16\pi} C^\alpha C_\alpha  \label{sourceL4} \end{equation}
We hypothesize that {\it in all cases}, the density of the source, $\rho_s$,  is given by (\ref{sourceL4}).  The conservative group in this case acts like the unitary group in quantum physics, since $C^\alpha C_\alpha$ is invariant under this group of transformations.  .   
This result lends support to the claim given above that the conservation group is the fundamental group for the quantum geometry of nature.  There is a family of manifolds, $\mathcal{Q}$, each of which is connected to the classical manifold, ${\cal M}_0$ via a conservative transformation.  

\section{Weak-field solution.}

We assume that there is a region where there are no sources and thus we have the trivial situation where we may choose $h^i_{\; \mu } = \delta^i_\mu$.     In the weak field situation, we perturb $h^i_{\;\mu}$ slightly via 
\begin{equation}
h^i_{\; \mu } \approx \; \delta^i_\mu + \frac12 H^i_{\; \mu} \label{wf1}
\end{equation}
where $H^i_{\; \mu}$ is a function of position with $\Bigl| H^i_{\;\mu} \Bigr| << 1 $.  Assuming that terms that are quadratic or higher in $H^i_{\;\mu}$ are negligible, we find that 
\begin{equation} 
g_{\mu\nu} \approx \; \eta_{\mu\nu} + \; H_{(\mu\nu)} \label{wf2}
\end{equation}
where $(\mu\nu)$ indicates symmetrization. It is easy to see also that
\begin{equation}
C_\mu \approx \; \frac12\Bigl(H^\alpha_{\;\mu ,\alpha} - H^\alpha_{\; \alpha , \mu}\Bigr) \label{wf3}
\end{equation}
To first order, the indices of terms containing $H^i_{\;\mu}$ are raised and lowered with $\eta^{\mu\nu}$ and $\eta_{\mu\nu}$.  We will also use harmonic coordinates ( \cite{Weinberg1972} , p. 254) which imply that $H_{(\alpha\mu)}^{\quad , \, \alpha} = \frac12 H^\alpha_{\; \alpha ,\mu}$.    In these coordinates, 
\begin{equation} 
C_\mu \approx \; -\frac12 H_{\mu\alpha}^{\;\;\; , \, \alpha} \qquad (\mathtt{in\;\; harmonic \;\; coords.}) \label{wf4}
\end{equation}
and \begin{equation}
C^\mu C_\mu \approx \; \frac14 H^{\mu \;\, , \alpha}_{\; \alpha} H_{\mu\beta}^{\;\;\; , \beta} \label{wf5} 
\end{equation}

The stress-energy tensor of the weak-field may be easily calculated.  We first note that the Ricci rotation coefficients are 
\begin{equation}  \gamma_{\nu\alpha\beta} \approx \; \frac12 \Bigl( H_{[\nu\alpha],\beta}+H_{\beta [\alpha , \nu]}+H_{[\alpha \beta , \nu]}    \Bigr)  \label{wf6}
\end{equation}
where $[\nu \alpha]$ indicates antisymmetrization of that pair of indices, for example $B_{[\nu \alpha]} = \frac12 \bigl( B_{\nu\alpha} - B_{\alpha\nu}\bigr)$.  The only terms that remain in $T_{\mu\nu} $  (see (\ref{e9})) are the terms:  $C_{\mu , \nu} - \eta_{\mu\nu}C^\alpha_{\; ;\alpha} + \gamma^{ \; \;\; \alpha}_{(\mu \;\;\nu) \, , \alpha}\;$  (see (\ref{e11}) and ({\ref{e12})).  We then find
\begin{equation}
8\pi T_{\mu\nu} \approx \; -\frac12 \Bigl( \;  H_{(\mu\nu)} - \frac12\eta_{\mu\nu} H^\alpha_{\;\;\alpha}    \;\Bigr)^{, \beta}_{\; , \beta} \label{wr7}
\end{equation}
For the {\it associated manifold}, we require $T_{\mu\nu} \approx 0$ to first order as is done in the standard theory\cite{MTW1973}.
Defining $\overline{H}_{\mu\nu} \equiv H_{(\mu\nu)}-\frac12\eta_{\mu\nu}H^\alpha_{\; \alpha}$, we see that this implies $\overline{H}_{\mu\nu \;\; , \beta}^{\;\;\; , \beta} = 0$ and thus gravitaional waves travel at the speed of light.

\section{Spherically Symmetric Solutions and a Framework for Galaxy Models.}

Spherically symmetric solutions have been the starting point for understanding the implications of general relativity.  The Schwarzschild metric and other spherically symmetric solutions are often used in situations where the symmetry is only approximate or indeed axial. 
 In spherical space-time coordinates ($\,(t,r,\theta, \phi)$ with $0\leq \theta \leq \pi \;$), an arbitrary spherically symmetric tetrad may be expressed by $h^i_{\;\; \mu} = $
\begin{equation} \left[ \begin{array}{cccc}
\; e^{\Phi(r)} & 0 & 0 & 0 \\
0 &\; e^{\Lambda(r)}\sin \theta \cos \phi \; & \; r\cos\theta\cos\phi \; & \; -r\sin\theta\sin\phi \; \\
0 & e^{\Lambda(r)}\sin\theta\sin\phi & r\cos\theta\sin\phi    & \; \;r\sin\theta\cos\phi \\
0 & e^{\Lambda(r)}\cos\theta\qquad & -r\sin\theta\qquad & 0
\end{array}
\right]
\label{e31} \end{equation}
where the upper index refers to the row. The curvature vector for this tetrad field is given by
\begin{equation} C_\mu = \frac{e^{\Lambda}}r \biggl[ \; 0,  \; 2 - e^{-\Lambda}\bigl(r\Phi^\prime + 2\bigr) , \; 0 , \;0  \biggr]  \label{e32} \end{equation}
where components are in the order $[t,r,\theta,\phi]$ and the prime denotes the derivative with respect to $r$. The tetrad (\ref{e31}) leads to the metric
\begin{equation}ds^2= -e^{2\Phi(r)} dt^2 +  e^{2\Lambda(r)}dr^2+r^2d\theta^2 +r^2\sin^2\theta d\phi^2 \, . \label{e33} \end{equation}
When $(r\Phi^\prime + 2)=2e^{\Lambda}$, then $C_\mu$ in equation (\ref{e32}) is identically zero and hence $(T_{\rm s})_{\mu\nu}  $ is identically zero leading to a free-field solution.

The metric (\ref{e33}) leads to a diagonal Einstein tensor with nonzero elements and thus the stress-energy tensor is:
\begin{eqnarray}8\pi T^t_{\; t}  = -8\pi \rho \, & \; =  \, \frac1{r^2}\bigl(-2re^{-2\Lambda}\Lambda^\prime + e^{-2\Lambda} - 1 \bigr)  \nonumber \\ & \; = -\frac2{r^2}\frac{d}{dr}\biggl[\frac12 r \Bigl(1  - e^{-2\Lambda}\Bigr) \biggr] \qquad  \label{e34} \end{eqnarray}
\begin{equation}8\pi T^r_{\; r} = \; 8\pi p_r \; = \; \frac1{r^2}\Bigl( 2re^{-2\Lambda}\Phi^\prime + e^{-2\Lambda}-1\Bigl)  \label{e35} \end{equation}
\begin{eqnarray} 8\pi T^\theta_{\; \theta}\, & \, =\; 8\pi T^\phi_{\;\, \phi}  \,= \; 8\pi p_T 
\qquad \qquad \qquad \qquad \qquad \quad \nonumber \\
  & \, = \, \frac{e^{-2\Lambda}}{r}\biggl(\, r\Phi^{\prime\prime}+r(\Phi^\prime)^2 - r\Phi^\prime \Lambda^\prime + \Phi^\prime - \Lambda^\prime \, \biggr) \;  \label{e36} \end{eqnarray}

We note that $T^t_{\; t}= - \rho$ depends only on $\Lambda(r)$.  Depending on whether $C_\mu$ is zero via (\ref{e32}) we will have a free-field or a field with sources.    The source term of the Lagrangian (\ref{e10}) that we will use is $L_s=\rho_s(r)$, where $\rho_s(r)$ is the density of the source as a function of $r$.  In the static case, from (\ref{e32}) and (\ref{sourceL4}), we find that 
\begin{equation}
  8\pi \rho_s = \, \frac12 C^\mu C_\mu \quad \Rightarrow \quad \rho_s = \frac1{16\pi} \, \biggl(\frac{\; 2\, e^\Lambda - 2 - r \Phi^\prime \;}{r e^\Lambda}  \biggr)^{\, 2}  \label{rhos}
\end{equation}
We assume that for realistic densities,  $\rho_s < \rho$ and we define $\rho_{dm} \equiv \rho - \rho_s$ to represent the density of dark matter.

\subsection{Particle motion in the spherically symmetric case.}

We exhibit here the general formulae governing such motion in a spherically symmetric solution of our field equations.  The equations of motion for a particle are based on adding an appropriate term to the Lagrangian.  We will use an approximate delta function, $\delta_\epsilon^4$ which is nonzero on a space-like volume equal to $\epsilon$ (the function used in the approximation is not important for our puposes).   Following \cite{dFC1990} we add a term
\begin{equation}  L_{p} = \rho_{p}(x)  = \mu \int \delta_\epsilon^4(x-\gamma(s))(-u^\mu u_\mu)^{\frac12} ds   \label{e67} \end{equation}
to the Lagrangian of (\ref{e10}). We assume $(\epsilon)^\frac13$ is small compared to the radial coordinate.  The mass of the particle will be denoted by $\mu$.  The path of the particle is represented by $\gamma(s)$ and its velocity is $u^\alpha=\frac{dx^\alpha}{d\tau}$.  We will use the "dot" notation for the components of $u^\alpha$, i.e. $u^\alpha = \langle \dot{t},\dot{r},\dot{\theta},\dot{\phi}\rangle$.  The condition  $T^{\beta \alpha}_{\; \;\;\; ; \beta }=0$ leads to (see \cite{dFC1990} and \cite{Green2020})
\begin{equation}  \; \frac{\mu \, u^\beta u^\alpha_{\; ;\beta} }{\sqrt{-u^\nu u_\nu}}  \, = \; \delta^\alpha_1 \, F_p \; \equiv \; \delta^\alpha_1 \, \epsilon \, e^{-2\Lambda} \biggl(-p_R^{\; \prime} + \frac2r \, \bigl(p_T-p_R\bigr)\,\biggr) \label{e68} \end{equation}
When $\alpha\neq 1$, (\ref{e68}) implies  $u^\beta u^\alpha_{\; ;\beta} = 0$ which is the usual geodesic equation.  Using (\ref{e35}) and (\ref{e36}) we find
\begin{equation}
  F_p \, = \; \frac{2\, \epsilon\, e^{-4\Lambda}\Phi^\prime (\Lambda^\prime + \Phi^\prime)\,}{\, r} \label{Fp}
\end{equation}

For orbital problems we use the standard approach of setting $\theta = \frac{\pi}2$.  For the remaining components we find:  
\begin{equation}   \ddot{\phi} + \frac2r \, \dot{r} \, \dot{\phi} \, = \, 0 \quad  \qquad  \Rightarrow \quad  \qquad  \dot{\phi} \; = \; \frac{L}{r^2} \label{phidot} \end{equation}
\begin{equation}  \ddot{t} \, + \, 2 \dot{\Phi} \, \dot{t} \, = \, 0 \quad \qquad \Rightarrow \quad \qquad \dot{t} \, = \, E \, e^{-2\Phi} \label{tdot} \end{equation}
and
\begin{equation}  \ddot{r} \, +  (\Phi^\prime + \Lambda^\prime )\, \dot{r}^2  + e^{-2\Lambda} \Phi^\prime + \frac{e^{-2\Lambda}}{r^3}  (r\Phi^\prime - 1) L^2 \, = \, \frac{1}{\mu} \, F_p   \label{rddot2} \end{equation}
where $L$ is a constant interpreted as the conserved angular momentum, $E$ is the constant energy and the $u^\alpha u_\alpha = -1 \; $   normalization has been used \cite{MTW1973}.  These results follow standard techniques and more details are given elsewhere \cite{Green2020a} .

\subsection{Infeasibility of The Free-Field Case}

In the free-field case, $C_\mu=0$ and hence the density of the source, $\rho_s$, is also identically zero.  The condition, $T^t_{\;t} = -  \rho $,  in the spherically symmetric case implies that $T^t_{\;t} = - \frac1{4\pi r^2} m^\prime (r) $ where the function $m(r)$ represents the total mass inside a ball of radius $r$.  Thus from (\ref{e34}) we see
\begin{equation}
  m(r) = \frac12 r \Bigl(1  - e^{-2\Lambda}\Bigr)
\end{equation}
(there is no constant added since clearly $m(0)=0$ as it should).  This implies that 
\begin{equation}
e^{2\Lambda} = \frac{1}{1-\frac{2m(r)}r}  \; \; , \label{e2Lambda}
\end{equation}  
hence $e^\Lambda = \Bigl(1-\frac{2m(r)}r\Bigr)^{-\frac12}$.  If $C_\mu = 0$, we find from (\ref{e32}) that $\Phi^\prime = \frac{2e^\Lambda}{r} - \frac2r\, $.  In the weak field, we assume that $m(r)\approx M$, a constant.  Itegrating we find that $ e^{2\Phi} = \frac1{16} \Bigl(1-\frac{M}r + \sqrt{1-\frac{2M}r}\Bigr)^4 \, \approx 1 - \frac{4M}r $.
But this implies that $g_{00} \approx -1 + \frac{4M}{r}$ which differs from the classical result of GR (which is $-1+\frac{2M}r$).  So we conclude that the free-field solution does not correspond to the classical solution or general relativity.

\subsection{Conditions for Classical Solutions}

For the model to be acceptable, we require that:\quad {\it i})  its Einstein tensor component $G^t_{\; t} = 8\pi T^t_{\; t}$ should match with $\, -\, 8\pi \rho\,$;\quad {\it ii}) in the weak-field scenario, its metric match with General Relativity's result in the $g_{tt}$ component; and \quad {\it iii}) its value of $\rho_s = \frac1{16\pi}  C^\mu C_\mu \, $ is such that $\, \rho_s < \rho\,$. Conditions {\it i}) and {\it ii}) imply that $C_\mu$  and hence $\rho_s$ are nonzero.

\subsection{Appropriate Choices for $C^\mu C_\mu$ and $\rho_s$}

In the spherically symmetric tetrad by using (\ref{e32}) and (\ref{rhos}),  we find that
\begin{equation}
  \Phi^\prime = \; \frac{2e^\Lambda}r - \frac2r-\beta(r) e^\Lambda  \label{Phiprime}
\end{equation}
where $\beta^2 = C^\mu C_\mu \, $ and hence $8\pi \rho_s = \frac12 \beta^2$.   Using a system of units where masses and distances are expressed in centimeters,  we see that $\beta^2$ must be dimensionally, cm$^{-2}$.    We will choose $\beta(r)$ using either the total density $\rho(r)$ or 
$\bar{\rho}(r) = \frac{m(r)}{r^3}$. Thus we choose to model $\beta(r)$ by 
\begin{equation}
\Bigl[\beta(r)\Bigr]^2 = \; 8 \pi \hat{\rho}(r) \, f\Bigl( \frac{m(r)}{r}\Bigr) \; , \quad , \label{betasquared}
\end{equation}
where $f(m/r)$ is a dimensionally neutral factor and either $\hat{\rho}(r) = \rho(r)$ or $\hat{\rho}(r) = \bar{\rho}(r)$.  
According the physical situation, we will employ either $\rho$ or $\bar{\rho}$ and choose the function $f(m/r)$ appropriately.

In the weak-field scenario, particle motion in the pure radial direction must satisfy $\ddot{r} \approx - \frac{m(r)}{r^2}$.   In this case $\frac{m(r)}{r}$ and $\beta(r)$ are very close to zero.  Thus $e^{\Lambda} = \Bigl(1- \frac{2m(r)}{r} \Bigr)^{-\frac12} \approx 1 + \frac{m(r)}{r}$ and from (\ref{Phiprime}) we find $\Phi^\prime \approx \frac2r \Bigl( 1+\frac{m(r)}r \Bigr) - \frac2r - \beta(r) $.  From (\ref{rddot2}) with $L=0$, $\dot{r}=0$ and assuming that $F_p \approx 0$,  we find that 
\begin{equation}
\ddot{r} \approx  - \frac{2m(r)}{r^2} + \beta(r)    \; .  \qquad  \mathtt{(pure \; radial\; motion)}  \label{rddot3}
\end{equation}
Thus in the weak-field scenario, $\beta(r) \approx \frac{m(r)}{r^2}$ and hence $8\pi \rho_s \approx \frac{m^2}{2r^4}$.

\subsection{A Framework for Modeling Bulge-dominated Galaxies }

We propose a galaxy mode framework consisting of three sections, a central {\it Bulge} (region dominated by baryonic matter), a {\it Mesosphere} (region dominated by dark matter) and an {\it Outside Region} where both dark matter and baryonic densities rapidly approach zero as $r$ increases.  Each region will have its own spherically symmetric tetrad solution and corresponding metric and stress-energy tensor (see Figure 1).    In contrast to the usual approach, we will model both the baryonic matter and the dark matter in a unified approach.  

In the {\it Bulge}, where $0\leq r \leq R_B$, let $M_B$ represent the total mass-enery which is mostly due to ordinary baryonic matter.  The fields are very strong and highly nontrivial in the {\it Bulge} and isothermal conditions are not expected.   Since the bulk of the visible matter is baryonic, we claim our galaxy model will apply to both spherical and axial galaxies.

\begin{figure*}[t]
\includegraphics[width=0.98\textwidth,height=0.16\textwidth]{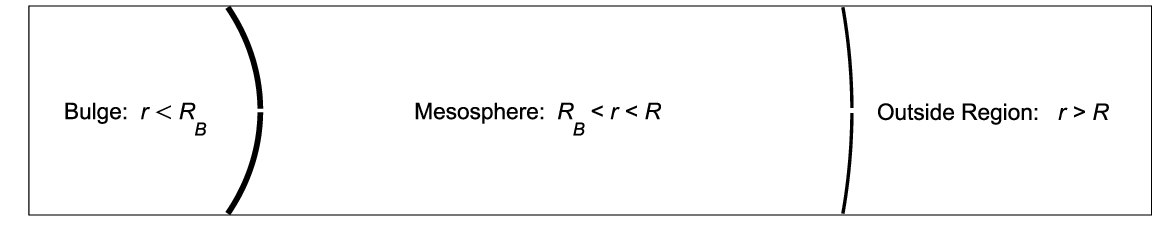}
\caption{Regions determined by the {\it Bulge} and {\it Mesosphere} in terms of the $r$-coordinate.}
\label{fig:1}
\end{figure*}

The interval $R_B < r \leq R$ corresponds to the {\it Mesosphere} where we assume that the density of the sources are small compared to the total density.  This implies that dark matter effects will appear in this region.  The weak-field approximation should apply, however and this will lead to a robust model.

For $r>R$, we have the {\it Outside Region} where densities are still nonzero, but rapidly approach zero as $r$ increases.  Starting with the {\it Bulge}, we will patch these solutions together to form a model that applies to the entire galaxy.  We will require that the metric tensor be continuous.  For the $g_{rr}$ component this is easily accomplished by making $m(r)$ continuous.  Once we find the {\it Outside Region} solution which matches the weak-field as $r\to\infty$, we will determine constants of integration for $\Phi$ (see (\ref{Phiprime})) at $r=R$ and then at $r=R_B$.    

There is great flexibility in this method for setting up a model, particularly in the modeling of the {\it Bulge}, the size and  extent of the {\it Mesosphere} and the particular source densiities that are used.   Numerical solutions may be utilized to more closely model actual galaxies, but our simplified models will nevertheless lead to several important results.

\section{A Plausible Bulge Model.}

In efforts to model the dark matter halo, one difficulty encountered was that these models often predict a central cusp where the density of dark matter diverges with order $\frac1r$ as $r\to 0\,$. Observations, however point to a roughly constant dark matter central density.  Various models for dark matter have attempted to address this issue (see \cite{Burkert2020}, \cite{Ghari2019}, \cite{Rodrigues2014} and \cite{Chavanis2019a}).   Nevertheless one must take care to avoid this ``cusp'' problem when setting up a model for the {\it Bulge}.   These models attempt to model dark matter separately from ordinary matter.  In our unified approach we see that dark matter must exist and contrary to being a special particle, it is a quantum geometrical effect of all matter.  In the spirit of ``geometry is gravity'' we claim that all matter affects the geometry in this previously unseen way.

A general model for the {\it Bulge} will not be our goal here.  But we will exhibit a plausible model which will show how more detailed models may be built.   Let $\alpha$ be a constant in units of cm$^{-1}$ and let $\kappa>0$ be a dimensionless constant. Consider a plausible model for the {\it Bulge} with $r \leq R_B$ corresponding to the following choices for $\rho$, $m(r)$ and $f\Bigl( \frac{m}r \Bigr)$:
\begin{equation}
\mathtt{Choose:}\;\;  8\pi \rho = \frac{4\alpha}r \;\;  \Leftrightarrow  \;\; m(r) = \alpha r^2 \label{bulge1}
\end{equation}
\begin{equation}
\mathtt{Choose:} \;  \; f\Bigl(\frac{m}{r}\Bigr) = \, 2 \Bigl[ \, 1 - \frac{\kappa m(r)}{r} \Bigr] \, = \,  2 \Bigl[ \, 1 - \kappa \alpha  r  \, \Bigr]   \label{bulge2}
\end{equation}
Thus using (\ref{betasquared}) with $\hat{\rho}=\rho$ we find 
\begin{equation}
  \beta^2 = \, C^\mu C_\mu \, = \, \frac{8\alpha}{r} \Bigl( 1 - \kappa \alpha r \Bigr)  \label{bulge3}
\end{equation}
Thus 
\begin{equation}
8\pi \rho_s  = \frac{4\alpha}r-4\kappa \alpha^2 \qquad \mathtt{and} \qquad 8\pi \rho_{dm} = 4\kappa \alpha^2  \label{bulge4}
\end{equation}
We note that while the density of the source diverges as $r\to 0$, the density of the dark matter is constant.  The dark matter mass function is $m_{dm} = \frac23 \kappa \alpha^2 r^3$.  These equations determine the spherically symmetric tetrad with
\begin{equation}
e^{\Lambda} = \frac1{\sqrt{1-2\alpha r}} \qquad  \mathtt{and} \qquad  e^{\Phi} = \frac{C e^{-Q(r)}}{(\frac12+\frac12\sqrt{1-2\alpha r})^4} \label{bulge5}
\end{equation}
where $Q(r) =  \int_0^r \frac{2\sqrt{2\alpha\,(\, 1-\kappa \alpha u\,)}}{\sqrt{\, u\, (1-2\alpha u \, )}} \, du $.  (If $\kappa=2$, $Q(r)=4\sqrt{2\alpha r}$.) The resulting metric is easily determined using (\ref{e33}).  The arbitrary constant $C$ should be chosen to make the metric continuous at $r=R_B$ with the {\it Mesosphere} metric.  Our {\it Bulge} model does restrict $R_B$ with the requirement: $\; R_B < \frac{1}{\kappa \alpha}$.  The resulting, complicated stress-energy tensor reveals that the temperature is nonconstant (see next section).

\section{Models for the Mesosphere. }

We denote the minimum value of $r$ at which the mass-energy density and the pressures are small (inner edge of {\it Mesosphere}) by $R_B$.     For $r > R_B $, we will use the ideal gas law with the average pressure to define the temperature per unit mass:  
\begin{equation}
  T = \frac{ \frac13 (\, p_r + \, 2\, p_T \,)}{\rho} \label{Tdef}
\end{equation}
Our hypothesis is that the {\it Mesosphere} is in thermodynamic equilibrium.    As pointed out by \cite{DSStern2014}, a thermodynamic principle may lead to an understanding of galaxy mass distributions and flat rotation curves.  

We now choose an appropriate expression for $\beta(r)$ as indicated in Section 3.4.  In the {\it Mesosphere}, we expect the density of the source to be small compared to the total density.  We note that $\frac{m(r)}{r} << 1$ in this region.   We expect the weak-field solution to apply within the {\it Mesosphere}, so from (\ref{rddot3}), we choose   
\begin{equation}
8\pi \rho_s = \frac{\bigl[m(r)\bigr]^2}{2 r^4} \qquad \mathtt{and} \qquad  \beta(r) = \frac{ m(r) }{\, r^2} \; .\label{betamodel}
\end{equation}
We see that the density of ordinary matter is very small compared to the average density at that point. For Condition {\it iii}) to be met, $8\pi \rho_{dm} =  \frac{ 2 \, m^\prime }{\, r^2} - \frac{ m^2}{2r^4} > 0$ which implies that $m^\prime > \frac{m^2}{4r^2}$.

From (\ref{betamodel}) and (\ref{Phiprime}) we find that 
\begin{equation} 
\Phi^\prime = \; \frac2r \Bigl( \, e^\Lambda  - 1 \, \Bigr) - \frac{  m}{r^2} e^\Lambda \;\; .
\end{equation}
Assuming $\frac{m}{r}$ is close to zero, we find that (to second order in $\frac{m}{r}$):
\begin{equation}
\Phi^\prime \approx \; \frac{m}{r^2} \, + \, \frac{2 m^2}{r^3}  \; \; . \label{Phiprime2}
\end{equation}
From (\ref{e2Lambda}) we find that 
\begin{equation}
\Lambda^\prime \approx  \; - \frac{m}{\, r^2} \, + \, \frac{\, m^\prime}{r} \, - \, \frac{2m^2}{\, r^3} \, + \, \frac{2mm^\prime}{r^2} \label{Lambdaprime2}
\end{equation}
Substituting these expressions into (\ref{e34}), (\ref{e35}) and (\ref{e36}) we find that
\begin{equation} 
  8\pi \rho = \; \frac{2m^\prime}{r^2} \qquad \mathtt{(exact)} \; , \label{e34a} 
\end{equation}
\begin{equation}  8\pi p_r \approx \; 0  \qquad \mathtt{(to\; second\; order)} \;\;  \mathtt{and}  \label{e35a} 
\end{equation}
\begin{equation} 8\pi p_T \approx \;   \frac{mm^\prime}{r^3} \qquad \mathtt{(to\; second\; order)}  \; .  \label{e36a}
\end{equation}
Using (\ref{Tdef}) we find the temperature is (to second order) is
\begin{equation}
T \; \approx \; \frac{m(r)}{3r} \;\; . \label{modelT}
\end{equation}
To second order in $\frac{m}r$, we find that $T$ is constant if   $\frac{m}r = \chi$, a constant.  This implies that $m(r)=\chi r$ and hence $m^\prime = \chi$.  Thus in this case we get (to second order)
\begin{equation} 
T \; \approx \; \frac{\chi}3   \;\; .  \label{T2}
\end{equation}

At the surface of the {\it Bulge}, the value, $m(R_B)=M_{B}$, represents nearly all of the baryonic mass of the galaxy.  Although $M_B$ for the entire galaxy is not completely determined by the value at $R_B$, we assume that the difference is small in relative terms.

Define a dimensionless constant $0<k<1$
\begin{equation}
  k-\frac12k^2 =\frac{\, M_B}{R_B} \; \quad \mathtt{with} \quad m(r) = \Bigl(\, k-\frac12k^2\Bigr)r \;    \label{halomass}
\end{equation}
(i.e. $\chi = k-\frac12 k^2$).  
This expression is slightly complicated, but will make the solution easier to analyze.   We note that when modeling a particular galaxy,  $k$ is not a free parameter because it is determined by $M_B$ and $R_B$ for that galaxy. However, the values of $R_B$ is usually not precisely known, so this gives some flexibility in setting up the model of the {\it Mesosphere}.    Thus at $r=R_B$,  we have
\begin{equation}
  e^{2\Lambda} \equiv \; \frac1{1-\frac{2m(r)}r} \; = \;  \frac1{(1-k)^2} \label{haloe2Lambda}
\end{equation}
and the $g_{rr}$ component of the metric is continuous at $r=R_B$.  

Using (\ref{betamodel}) we find 
\begin{equation}
  \beta(r) \, =  \; \frac{k-\frac12k^2}{r} \; \;  \quad \Rightarrow \quad \; 8\pi \rho_s = \; \frac{ k^2(1-\frac12 k)^2}{2r^2}  \label{halobeta}
\end{equation}
and  $C^\mu C_\mu =  \, \frac{k^2(1-\frac12k)^2}{r^2}$.  From (\ref{Phiprime2}) we find
\begin{equation}
  \Phi^\prime(r) \; = \; \frac{k+\frac12k^2}{(1-k)r} \; \quad \Rightarrow \quad \; e^{2\Phi} \; = \; \biggl(\; \frac{r}{R_0} \;\biggr)^{\frac{2k + k^2}{1-k}} \label{haloe2Phi}
\end{equation}
where $R_0$ is the constant of integration which will be determined below.  Thus, in the {\it Mesosphere}, we have the following metric:
\begin{equation}
  ds^2 = - \Bigl( \frac{r}{R_0} \Bigr)^{\frac{2k +k^2}{1-k}} dt^2 +  \frac1{(1-k)^2}dr^2 +  r^2(d\theta^2 + \sin^2\theta d\phi^2)  \label{halometric}
\end{equation}
for $R_B \leq \; r \; \leq R \, $.  Since there are no singularities in this metric (\ref{halometric}), we find no restriction on the value of $R$, i.e., the size of the {\it Mesosphere} is an adjustable parameter.  
From (\ref{e34}),(\ref{e35}) and (\ref{e36}) we find
\begin{equation}
  8\pi \rho \; =  \; \frac{k(2-k)}{r^2} \label{halodens}
\end{equation}
\begin{equation}
  8\pi p_r \; = \; \frac{-k^3}{r^2} \label{halopr}
\end{equation}
\begin{equation}
  8\pi p_T \; = \; \frac{k^2\Bigl[ 2+k \Bigr]^2}{4r^2} \label{halopT}
\end{equation}
where in constrast to (\ref{e35a}) and (\ref{e36a}), these equations are now exact.  
Using (\ref{Tdef}), we find there is a (exactly) constant temperature per unit mass which is given by 
\begin{equation}
T= \frac{\, k(4+2k+k^2)}{6(2-k)} \; \approx \frac{k+k^2}{3} \; \; .  \label{Case1Temp}
\end{equation}
From the definition of $k$, we estimate that $k$ is small (estimates are $10^{-8} < k < 10^{-4}$), so terms involving $k^2$  and $k^3$ may be neglected in many calculations.

The density of dark matter is found to be 
\begin{equation}
8\pi \rho_{dm}   = \; \frac{k(16-12k+4k^2-k^3)}{8r^2} \; \approx \frac{k(2-\frac32 k)}{r^2}  \;\; .
\label{rhodm1}
\end{equation}
From (\ref{halobeta}) and (\ref{halodens}) we find $\frac{\rho_s}{\rho} = \frac{k(2-k)}8\approx \frac{k}{8}$.
Since $k$ is typically very small, then nearly all the additional mass-energy in the {\it Mesosphere} is attributable to dark matter.

\section{Example of How to Model the Outside Region.  Stitching the Regions together.}

There is a great deal of flexibility in the model for the {\it Outside Region} where $r>\, R\, $.   We expect the weak-filed solution to apply and also the solution should be approximately isothermal.   As for the {\it Bulge} solution we will not derive a general solution, but will exhibit a plausible mode that shows how to model this region.  For $r>R$, we have several possible solutions that meet our conditions. 

In our example, we will utilize a model \cite{Green2020a} that is consistent with the external Schwarzschild solution of general relativity, but differs in that the Einstein tensor is again required to be nonzero because of our conditions for an acceptable model.   The solution is approximately isothermal, but with a negative temperature.    Recall $\lim_{r\to\infty} m(r) = M$.   One algebraically simple solution used in our solar system model \cite{Green2020a} is $m(r) = M(1- \frac{M}{2r})$.   In this case we recall from (\ref{rddot3}) that to leading order, $\beta(r) = \frac{m(r)}{r^2}$.  We are free to choose the nonleading terms for our convenience.    For these reasons we choose $\beta(r) = \frac{M}{r^2(1+\frac{M}r)}\;$ and thus $\, C_\mu C^\mu = \frac{M^2}{r^4(1+\frac{M}r)^2}$.   Using (\ref{Phiprime}) we solve for $\Phi(r)$.  The resulting metric is
\begin{equation}
  ds^2 = \, - \Bigl(1-\frac{M}r \Bigr)^3\Bigl(1+\frac{M}r \Bigr) \, dt^2 + \frac{1}{(1-\frac{M}r)^2}\, dr^2 + r^2 d\Omega^2  \label{outmetric}
\end{equation}
where $ d\Omega^2 = d\theta^2 + \sin^2\theta d\phi^2   $.   The relevant functions are
\begin{eqnarray}
  \; 8\pi \rho \; = \; \frac{M^2}{r^4} \qquad \qquad & \qquad   8\pi p_T = \; - \frac{M^2(1-\frac{5M}{r}-\frac{5M^2}{r^2})}{r^4(1+\frac{M}{r})^2}    \nonumber \\
   m(r) \; = \; M  \, - \, \frac{M^2}{2r} \quad & \quad 8\pi p_r = \frac{M^2(1-\frac{3M}{r})}{r^4 (1+\frac{M}{r} )} \qquad \;\; \label{outfunctions} \\
  8\pi \rho_s = \, \frac{M^2}{2r^4\Bigl(1+\frac{M}r\Bigr)^2}   & \qquad  m_s \approx \frac{M^2}4\biggl(\frac1R-\frac1r\biggr) + M_B   \nonumber
\end{eqnarray}

We require that the $m(r)$ value must agree at $r=R$ for the {\it Mesosphere} and the {\it Outside Region} models.  From (\ref{halomass}) and (\ref{outfunctions})  we then see that $\, M(1-\frac{M}{2R})\, = \, k(1-\frac12 k)R\,$ 
and thus we conclude that $\, M \, = \, kR\,$.  Thus,
\begin{equation}
  k =  \frac{M}R   \quad  \Rightarrow   \quad  \frac{M}{R}\Bigl(1-\frac{M}{2R}\Bigr) = \frac{M_B}{R_B}  \quad  \Rightarrow \quad  \frac{M}{R}\; \approx \; \frac{M_B}{R_B} \,  .  \label{ratios}
\end{equation}
Hence $\frac{M}{M_B}\approx \frac{R}{R_B}$.

We note that 
\begin{equation}
\frac{M}{M_B}  \approx \frac{R}{R_B}  \quad .   \label{ratiorule}
\end{equation}
Thus the ratio of the total mass-energy to the {\it Bulge} (baryonic) mass-energy is related to the ratio of the radius of the outer edge of the {\it Mesosphere} to the radius of the {\it Bulge}.

\subsection{Determining Constants of Integration.}  We now determine the value of $R_0$ which is a constant of integration in the solution for the metric in the {\it Mesosphere}, $R_B < r < R$ coming from solving for $\Phi$.  The value will be determined by matching the $g_{tt}$ component of the metrics (\ref{halometric}) and (\ref{outmetric}) at $r=R$. From (\ref{ratios}),  $k=\frac{M}{R}$ and so from (\ref{outmetric}) at $r=R$, $e^{2\Phi} = (1-k)^3(1+k)$.   Thus we find that 
\begin{equation} \biggl(\frac{R}{R_0}\biggr)^{\frac{\,2k+k^2}{1-k}} = \; (1-k)^3(1+k) \;\; . 
\end{equation} 
Hence 
\begin{equation}
\frac{R_0}{R} = \biggl[  \;   (1-k)^3(1+k) \; \biggr]^{\frac{-(1-k)}{\,2k+k^2}} \; . \label{R0eq}
\end{equation}
Since $k$ is small, we use the limit as $k$ goes to zero to approximate the right hand side of (\ref{R0eq}).  The result is
\begin{equation}
R_0 = \; R\, e^1 \; . 
\end{equation}
Hence the metric for the {\it Mesosphere} is
\begin{equation}
  ds^2 = \, - \, \biggl(\frac{r}{eR}\biggr)^{\, \frac{2k+k^2}{1-k}} dt^2 + \frac{dr^2}{(1-k)^2} + r^2 d\theta^2 + r^2 \sin^2\theta d\phi^2 \; . \label{halometric2}
\end{equation}
For our {\it Bulge} model example, we also determine the arbitrary constant, $C$.  At $r=R_B$ using (\ref{bulge5}), we find that 
\begin{equation}
C \; = \; \Bigl(\frac12 + \frac12\sqrt{1-2\alpha R_B}\Bigr)^4 \biggl(\frac{R_B}{eR}\biggr)^{\, \frac{2k+k^2}{2(1-k)}} e^{Q(R_B)}  \; .
\end{equation}

\section{Motion of Test Bodies within the Mesosphere.}

In the {\it Mesosphere}, $R_B < r < R$, we have the metric (\ref{halometric2}).  Using (\ref{Fp}), we find that a very small outward force on a test body in this region is  given by
\begin{equation}
  F_p \, \approx \; \frac{2 k^2}{\hat{\rho}\, r^3} \; \; , \label{Fp2}
\end{equation}
where $\hat{\rho}$ is the density of the test body.  As $k^2$ is very small and $r$ is very large, we will ignore this term in what follows.  
Thus, we find from (\ref{rddot2}):
\begin{equation}
  \ddot{r} +\frac{k+\frac32 k^2}{r} \dot{r}^2 + \frac{k -\frac12 k^2}{r} \; - \;  \frac{\Bigl[ 1-3k+\frac32 k^2 \Bigr] L^2}{r^3}  \approx  0  \; .  \label{halorddot}
\end{equation}

For pure radial motion($L=0$) in the {\it Mesosphere} ( with $R_B< r < R$  )  when velocities are small, (\ref{halorddot}) yields
\begin{equation}
\ddot{r} \approx \; -\, \frac{k-\frac12 k^2}{r}   \; = \; - \frac{m(r)}{r^2}  \label{halopureradial}
\end{equation}
as expected.

\subsection{Flat Rotation Velocity Curves within the Mesosphere.}

For circular orbits we set both $\dot{r}$ and $\ddot{r}$ to zero in (\ref{halorddot}).    Replacing $L$ with $r^2 \dot{\phi}$, we find
\begin{equation}
  r^2  \dot{\phi}^2  = \; \frac{k-\frac12k^2}{1-3k+\frac32k^2} \; \approx \; k + \frac52k^2 \label{dphidt}
\end{equation}
We denote the circular velocity by $v_r$ which is given by $v_r = r\frac{d\phi}{dt} $.  Assuming velocities are small and using the normalization $u^\nu u_\nu = -1$, we find that $-e^{2\Phi} \dot{t}^2 \approx -1 $ and hence
\begin{equation}
\dot{t}^2 \approx \Bigl(\frac{r}{eR}\Bigr)^{-2k-3k^2}
\end{equation}
Thus $\Bigl(\frac{d\phi}{dt}\Bigr)^2=\Bigl(\frac{\dot{\phi}}{\dot{t}} \Bigr)^2 = \bigl(\dot{\phi}^2 \bigr) \Bigl( \frac{r}{eR} \Bigr)^{2k+3k^2}$.  Define $\xi$ as the ratio of the outer radius of the {\it Mesosphere} to the radius of the {\it Bulge}:  $\; \xi \equiv \frac{R}{R_B}$.  Then for $R_B<r<R$ we find
\begin{eqnarray}
1 > \Bigl(\frac{r}{eR} \Bigr)^{2k+3k^2} &\, = \Bigl( \frac{r}{e \xi R_B} \Bigr)^{2k+3k^2}  > \Bigl(\frac{1}{e \xi}\Bigr)^{2k+3k^2} \nonumber \\  & > \, 1 - (2k+3k^2)\ln(e \xi ) \, ,  \qquad \;\;
\end{eqnarray}
where the last expression is due to the Taylor expansion of an exponential. Thus
\begin{equation} 
r^2 \dot{\phi}^2  > r^2 \dot{\phi}^2 \Bigl( \frac{r}{eR} \Bigr)^{2k+3k^2} > r^2 \dot{\phi}^2 \Bigl[ 1 - (2k+3k^2)\ln(e \xi ) \Bigr]
\end{equation}  
Using (\ref{dphidt}) and dropping terms of order $k^3$ or higher,  we find
\begin{equation}
k + \frac52 k^2 > \; v_r^{\; 2} > \;  k + \frac52 k^2 - 2\ln( e \xi ) k^2
\end{equation}
for $R_B < r \leq R$.  
To first order,
\begin{equation}
(v_r)^2 \approx k \qquad , \mathtt{ \rm \; i.e.} \quad  v_r \approx \sqrt{k} \quad , \;\; \mathtt{\rm for } \quad R_B < r \leq R \label{e84} 
\end{equation}
In conventional units, $v_r \approx 2.998 \times 10^5 \sqrt{k} $ km s$^{-1}$.  For example if $k = 10^{-7}$ (fairly typical for a galaxy) then $v_r\approx  95 \, $km/s.  This constant velocity curve would extend from the disk out to outer edge of the {\it Mesosphere}.  So we conclude that our theory along with the demand for an isothermal solution leads to a flat rotation velocity curve. 

\subsection{Connection to the Tully-Fisher relation and the Radial Acceleration Relation.  }

The baryonic Tully-Fisher relation \cite{Tully1977} states that $M_B \propto (v_r)^4 $. Also, McGaugh, Lelli and Schombert find from observational data from 153 galaxies that $\frac{g_{bar}}{g_{obs}} \approx \, 1-e^{-\sqrt{g_{bar}/g_\dagger}} \,$,  where $g_{bar}$ is the inferred acceleration due to the baryonic matter and $g_{obs}$ is the actual acceleration due to the observed rotation curve \cite{McGaugh2016}.   This relation, called the radial acceleration relation (RAR), may also be expressed \cite{McGaugh2016} in the form $\frac{g_{dm}}{g_{bar}} = \frac1{e^{\sqrt{g_{bar}/g_\dagger}} -\, 1 \,}$.     Given the value of $\frac{g_{dm}}{g_{bar}}$ at $r=R_B$, then we use the RAR  to find  
\begin{equation}
g_{bar} = \Bigl[ \ln(1 + \frac{g_{bar}}{g_{dm}})\Bigr]^2 g_\dagger \qquad \mathtt{(\, at \;\, }r = R_B ) .  \label{gbar} 
\end{equation}   

\subsubsection{Example.}  A rough estimate of the $r$-value where the {\it Mesosphere} begins would be the point at which the accelereation due to dark matter grows to a value of $p$, where roughly $0.01 < p < 0.50$ of the baryonic matter (i.e. between 1\% and 50\%).  Thus, $g_{bar} = (\ln p)^2 g_\dagger $.  Since the weak-field solution applies, we estimate that $\frac{M_B}{R_B^{\; 2}} \approx (\ln p)^2 g_\dagger$.   For example, if $p=0.01$,     $\frac{M_B}{R_B^{\; 2}} \approx 21.3g_\dagger$  From (\ref{halomass}) and (\ref{e84}) we see that
\begin{equation}
v_r^{\; 4} \approx k^2 \approx \biggl(\frac{M_B}{R_B}\biggr)^2 \approx (\ln p)^2 g_{\dagger} M_B \; .  \label{TFequation}
\end{equation}
Combining our results with the results of \cite{McGaugh2016} implies our example satisfies the baryonic Tully-Fisher relation with a constant of proportionality of $\frac1{(\ln p)^2 g_\dagger }$.  For example, if $p=0.01$, $M_B \approx \frac{1}{21.3 g_\dagger} v_r^{\; 4}$.  

\subsection{Procedure for developing a galaxy model for a rotationally supported galaxy using RAR and using our Example Bulge Solution.}

\begin{table*}
\caption{Galaxy model examples based on choices of $\frac{g_{dm}}{g_{bar}}$ and $R_B$.}
\label{tab1}    
\begin{tabular}{lllllll}
\hline\noalign{\smallskip}
 $ \frac{g_{dm}}{g_{bar}} $  &   $R_B$(kpc)   &   $\alpha$ (cm$^{-1}$)  &   \quad $ k $  & $ v_r$(km/s)  & $\kappa$ &  $ \rho_{dm} $(cm$^{-2}$)  \\
\noalign{\smallskip}\hline\noalign{\smallskip}
$0.030$ &  $\;\; 1.0$  & $1.74 \times 10^{-28}$ & $5.37\times 10^{-7}$ & $\;219.6$ & 83867& $4.04\times 10^{-52}$ \\
$0.200$ & $\;\; 1.0 $ & $  5.20\times 10^{-29} $ & $ 1.61\times 10^{-7} $ & $\;120.1$ &1869180& $8.05\times 10^{-52}$ \\
$0.400$ & $\;\; 1.0$ & $ 2.97 \times 10^{-29} $ & $ 9.15\times 10^{-8} $ & $\;\;\;90.7$ & 6554740& $9.18\times 10^{-52}$ \\
$0.050$ & $\;\; 2.0 $& $ 1.31\times 10^{-28} $ & $ 8.11\times 10^{-7} $ & $\;270.0$ & 92486& $2.54\times 10^{-52}$ \\
$0.200$ & $\;\; 2.0 $& $ 5.20\times 10^{-28} $ & $ 3.21\times 10^{-7} $ & $\;169.9$ & 934590& $4.02\times 10^{-52}$ \\
$0.200$ & $\;\; 5.0 $& $ 5.20\times 10^{-29} $ & $ 8.03\times 10^{-7} $ & $\;268.6$ & 373836& $1.61\times 10^{-52}$ \\
$0.500$ & $ 10.0$& $2.44\times 10^{-29} $ & $7.54\times 10^{-7} $ & $\;260.4$ & 994378&  $9.45\times 10^{-53}$ \\
\noalign{\smallskip}\hline
\end{tabular}
\end{table*}

\quad $\,$ 1)  Input the value of the ratio, $\frac{g_{dm}}{g_{bar}}$  and the value of $R_B$.  Use (\ref{gbar}) to find $g_{bar}$ and the (total) acceleration:  $\; g_{tot}\, = \, g_{bar}\Bigl(\, 1 \, + \, \frac{g_{dm}}{g_{bar}}\Bigr)$ at $r=R_B$.   

2) Using (\ref{bulge1}), we find the total acceleration at the boundary of the {\it Bulge} to be $\frac{m(r)}{R_B^{\,2}} = \alpha$.    

3)  Find $k = \frac{M_B}{R_B} = \alpha R_B$.   

4)  Determine the rotation velocity:  $v_r \approx \sqrt{k}$.  

5)  From the ratio of the mass functions for dark matter and baryonic matter, we also determine $\kappa =  \frac{3}{2k} \cdot \frac{g_{dm}}{g_{bar}}$.   

6)  For comparison purposes to standard dark matter halo models (see \cite{Donato2009}, \cite{Ghari2019}, \cite{Rodrigues2014} and  \cite{Burkert2020}), find the central density of dark matter: $\rho_{dm} = \frac{\kappa \alpha^2}{2\pi}$ (compare to $\rho_0$ in dm profiles).  

In Table \ref{tab1} we list several examples for hypothetical galaxies.  We note that the values of $v_r$ and $\rho_{dm}$ are reasonable values based on current observations.

\subsection{Motion of Test Bodies in the Outside Region.}
Using (\ref{outmetric}) and (\ref{outfunctions}) in the equations of motion for the pure radial case with low velocities yields
\begin{equation}
  \ddot{r} \; \approx \; - \, \frac{M}{r^2}\biggl(\, 1 - \frac{2M^2}{r^2} \, \biggr) \label{rddotout}
\end{equation}
and from (\ref{dphidt}) we find
\begin{equation}
  (v_r)^2 \; \approx \; \frac{M}{r}\biggl( \, 1 + \frac{3M}r \, \biggr) \label{vrout}
\end{equation}
At $r=R$, $\frac{M}R\approx k$.  Thus, in the {\it Outside Region}, the rotation velocity curve gradually drops from the {\it Mesosphere} value approaching zero as $r\to \infty$ as expected.

\section{Conclusion.}

The theory based on the conservative transformation group leads to a spherically symmetric solution for the free field, which must have nonzero densities and pressures.  However, we find that this free-field equation is not acceptable and thus introduce a density of the source term to the Lagrangian.  We claim that bulge-dominated galaxies naturally lead to a spherically symmetric framework consisting of three regions:  the {\it Bulge}, the {\it Mesosphere} and the {\it Outside Region}.  

Assuming the density of dark matter is not cuspy (i.e., bounded) inside the {\it Bulge}, we show how a reasonable model may be set up for the {\it Bulge}.   Our example solution for this central core region is one of many possible solutions. For the {\it Mesosphere}, we then find an acceptable spherically symmetric solution that is isothermal and meets the conditions of the weak-field solution in GR.  The {\it Mesosphere} solution is also dominated by dark matter.   We find an {\it Outside Region} solution that is reasonable, also satisfies the weak-field condition, but is only approximately isothermal.  We combine these solutions to get a continuous model (for the tetrad and metric).  

The analysis of the motion of a test body shows that our model implies there is a flat rotation curve, determined by the parameter $k$.  Using the RAR, we are able to set up reasonable galaxy models for rotationally supported galaxies.  These models were shown to be determined by the radius of the {\it Bulge}, $R_B$ and the ratio of the accelerations, $\frac{g_{dm}}{g_{bar}}$ evaluated at $r=R_B$.  We note that the theory does not theoretically restrict the size of the {\it Mesosphere}.  Our framework reasonably models galaxies that are bulge-dominated, but also gives insights into what more accurate models based on the conservative transformation group may predict.  Further progress in developing isothermal models that more closely model disk-dominated spirals and dwarf galaxies may strengthen these results.   

In all cases that have been considered, the extension of the covariance group to the conservative transformation group leads to  automatic inclusion of dark matter and dark energy.  As our theory suggests along with many recent results, dark matter and baryonic matter are not independent of one another.  It appears that dark matter is simply a geometrical effect of baryonic matter, but much work is needed to fully develop this theory.  If the conservation group is found to lead to the geometry of the quantum, dark matter and dark energy may be interpreted as quantum effects due to this geometry.

{\bf Data Availability Statement:}  No Data associated in the manuscript. 

\bibliography{ELGResearch}

@article{Burkert2020,
  author={Burkert, A},
  journal={Astrophys J},
  volume={904},
  pages={161},
  year={2020}
}

@article{Chavanis2019a,
  author = {Chavanis, Pierre-Henri},
  journal = {Phys Rev D},
  volume = {100},
  issue = {12},
  pages = {123506},
  numpages = {41},
  year = {2019}
}

@article{Donato2009,
  author={Donato, F et al},
  journal={Mon Not R Astron Soc},
  volume={397},
  pages={1169},
  year={2009}
}

@article{DSStern2014,
    author = {Davidson, J and Sarker, S K and Stern, A },
    journal = {Astrophys J},
    volume = {788},
    pages = {37},
    year = {2014}
}

@article{Ghari2019,
    author={Ghari, A and Famaey, B and Laporte, C and Haghi, H},
    journal={Astron  Astrophys},
    volume={623},
    pages={23},
    year={2019}
}

@article{Green2009,
    author = {Green, E L },
    journal = {Int J Theor Phys},
    volume = {48},
    pages = {323-336},
    year = {2009}
}

@article{Green2020,
    author = {Green, E L },
    journal = {Electron J Theor Phys},
    volume = {38},
    pages = {(to appear)},
    year = {2021}
}

@article{Green2020a,
   author= {Green, E L },
   journal = {Gen Rel Grav},
   volume = {52},
   number = {68},
   year = {2020}
}

@article{McGaugh2016,
    author = {McGaugh, S and Lelli, F and Schombert, J},
    journal = {Phys Rev Lett},
    volume = {117},
    pages = {201101},
    year = {2016}
}

@article{Pandres1981,
    author = {Pandres, D },
    journal = {Phys Rev D},
    volume = {24},
    pages = {1499-1508},
    year = {1981}
}

@article{Pandres1984,
    author = {Pandres, D },
    journal = {Phys Rev D},
    volume = {30},
    pages = {317-324},
    year = {1984}
}

@article{Pandres2009,
    author = {Pandres, D },
    journal = {Gen Rel Grav},
    volume = {41},
    pages = {2501-2528},
    year = {2009}
}

@article{PG2003,
    author = {Pandres, D and Green, E L },
    journal = {Int J Theor Phys},
    volume = {42},
    pages = {1849-1873},
    year = {2003}
}

@article{Rodrigues2014,
  author={Rodrigues, D and de Oliveira, P and Fabris J and Gentile, G},
  journal={Mon Not R Astron Soc},
  volume={445},
  pages={3823-3838},
  year={2014}
}

@article{Tully1977,
    author = {Tully, R B and Fisher, J R },
    journal = {Astron Astrophys},
    volume = {54},
    pages = {661},
    year = {1977}
}

@book{Einstein1949,
    author = {Einstein, A},
    title={Albert Einstein Philosopher-Scientist},
    year={1949},
    publisher={Harper},
    address = {New York}
}

@book{MTW1973,
    author = {Misner, C and Thorne, K and Wheeler, J A },
    title = {Gravitation},
    year = {1973},
    publisher = {W. H. Freeman and Company},
    address = {New York}
}

@book{Schouten1954,
    author = {Schouten, J A  },
    title = {Ricci-Calculus},
    edition = {2},
    year = {1954},
    publisher = {North Holland},
    address = {Amsterdam}
}

@book{Schrodinger1950,
    author = {Schrodinger, E  },
    title = {Space-Time Structure},
    year = {1950},
    publisher = {Cambridge University Press},
    address = {Cambridge}
}

@book{Weinberg1972,
    author = {Weinberg, S },
    title = {Gravitation and Cosmology},
    year = {1972},
    publisher = {Wiley},
    address = {New York}
}

@book{dFC1990,
    author = {de Felice, F and Clarke, C J S },
    title = {Relativity on Curved Manifolds},
    year = {1990},
    publisher = {Cambridge University Press},
    address = {Cambridge}
}

\end{document}